\documentclass[preprint2]{aastex}

\begin{document}

\title{Transit flow models for low and high mass protostars}

\author{C. Combet\altaffilmark{1,2}, T. Lery\altaffilmark{1}, 
G.C.Murphy\altaffilmark{1,3}}
\altaffiltext{1}{Dublin Institute for Advanced Studies, 
5 Merrion Square, Dublin 2, Ireland; combet@cp.dias.ie}
\altaffiltext{2}{Laboratoire de l'Univers et de ses Th\'eories, 
5 Place J. Janssen, B\^at 18 LAM, 92190 Meudon, France}
\altaffiltext{3}{Physics Department, Trinity College Dublin, Dublin 2, Ireland}

\shorttitle{Transit flow models for YSO}
\shortauthors{Combet et al.}

\begin{abstract}

In this work, the gas infall and the formation of outflows 
around low and high mass protostars are investigated. 
A radial self-similar approach to model
the transit of the molecular gas around the central object is employed.
We include gravitational and radiative fields to produce  
heated pressure-driven outflows with 
magneto-centrifugal acceleration and collimation.
Outflow solutions with negligible or 
vanishing magnetic field are reported. They indicate that thermodynamics
is a sufficient engine to generate an outflow.
The magnetized solutions show dynamically significant 
differences in the axial region, precisely where the radial velocity and 
collimation are the largest. 
They compare quantitatively well with observations.
The influence of the opacity on the transit solutions is also studied. 
It is found that, when dust is not the dominant coolant, such as in
the primordial universe, mass infall rates have substantial larger values
in the equatorial region. This suggests that 
star forming in a dust-free environment should be
able to accrete much more mass and become more massive than present 
day protostars. 
It is also suggested that molecular outflows may be dominated by the
global transit of material around the protostar during the very early stages 
of star formation, especially in the case of massive or 
dust-free star formation. 

\end{abstract}


\section{Introduction}

The understanding of star formation has changed dramatically during
the last two decades thanks to the development of both instrumental
and numerical techniques. However, while the 
picture of the main stages of low mass star formation is becoming clearer, 
the situation for
massive objects remains problematic. In the present paper, we address
the problem of the origin of outflows and infall around protostars over a wide 
range of masses.

At present, the stages of low mass protostar evolution are empirically 
divided in 
four classes (see for example \citet{2000prpl.conf...59A} for a review)
corresponding to the evolutionary sequence from the initial collapse
to a new star.
``Class 0'' protostars correspond to the earliest stage of star 
formation. Observationally, they appear deeply embedded in a circumstellar
dusty envelope (detected in sub-millimeter wavelengths) that is more massive
than the central stellar mass. Furthermore, they show powerful bipolar 
ejections of material in the form of collimated CO outflows which 
distinguish them from the pre-stellar phase of star formation 
(a gravitationally bound core within a molecular cloud).

The next stage of the star formation process corresponds to
``Class I'' objects which are typically $10^5$ year old and are characterized
by a positive slope of their spectral energy distributions (SEDs) between
2.2 and 10-25 $\mu$m. They are still surrounded by a diffuse circumstellar
envelope but are also surrounded by a disk from which the accretion 
onto the central object takes place. At this stage, the mass contained in the 
envelope is smaller than the mass of the central protostar.
Like Class 0 objects, they also produce bipolar 
outflows (optical jets and molecular outflows) but these are less
($\sim$ 1 order of magnitude) powerful than those observed for 
Class 0 protostars \citep{1996A&A...311..858B}.

Class II and III objects (resp. classical and weak-line T Tauri stars)
are the final two stages in evolutionary 
sequence of protostars. These correspond to pre-main sequence stars
surrounded by an accretion disk (optically thick (Classical) or thin depending
on the degree of evolution) but which do not have a circumstellar
envelope, as opposed to Class 0 or 1 objects.

Formation of massive stars ($M \ge 10 M_\odot$), is 
less well understood but two main theories have emerged:
i) the \emph{coalescence scenario} in which massive stars are formed by
the merging of intermediate mass objects \citep{1998MNRAS.298...93B} and 
ii) the \emph{accretion scenario}, originally applied to massive star by 
\citet{1994ApJS...95..517B},
where a massive star forms thanks to large accretion rates 
($10^{-4}-10^{-2} \; M_\odot$) onto the central object 
\citep{2000A&A...359.1025N,2001A&A...373..190B}.
The main difficulty of the merging scenario arises when one considers
the high stellar densities needed for it to be efficient.
On the other hand, it is difficult to accrete gas onto a very 
luminous star so that the upper mass of the accretion scenario is
rapidly reached.

Whatever the mass of the central object is, it appears that 
the accretion of material is accompanied by strong bipolar outflows.
During the last two decades, starting with the first outflow observation 
from a forming star by \citet{1980ApJ...239L..17S}, the number of
outflow observations has increased to the point that  
this phenomenon is now widely believed to affect every forming star.
Bipolar outflows allow the forming star to transport 
any excess of angular momentum. They can be classified as either
atomic jets or molecular outflows, according to their properties.

Jets are fast 
($\sim 100-300$~km~s$^{-1}$), 
well-collimated (opening angle~$<$ 10$^\circ$), 
and mostly constituted of atomic material.
Many models  and simulations attempt to describe
how they can be launched from the magnetized accretion disk of the protostar 
\citep{1997A&A...319..340F,1997ApJ...482..712O}, 
and their magneto-centrifugal
origin is commonly accepted. However, the nature of the exact mechanism,
be it a disk wind \citep{1982MNRAS.199..883B} or the asymptotic
collimation of an $X$-wind  
\citep{1995ApJ...455L.155S,2000prpl.conf..789S}, is still debated.

Molecular outflows (often traced
by CO and H$_2$ molecules) 
are more massive, slower ($\sim$ tens of km~s$^{-1}$) and less collimated 
than jets. The driving mechanism of these outflows is still open to discussion
 but the jet-driven bow shock model 
\citep{1993A&A...278..267R, 1994ApJ...426..204C, 1999A&A...345..977D,
2001ApJ...557..443O} and the 
wind-driven model \citep{1991fse..conf...23S,2000prpl.conf..789S} 
are the two main theories that have emerged. In the jet-driven
model, the bow shock surface created at the head of the jet transfers 
momentum to the ambient medium thus producing a thin shell that is
identified as the molecular outflow. In the second approach,
a wide-angle wind creates two bipolar wind-blown bubbles that 
sweep up the ambient material (then referred to as the molecular outflow).

For low and intermediate mass star formation, the morphologies and
kinematics of the observed outflows are generally explained by
one of these two models: some outflows present jet-driven features 
(e.g. HH212, Orion S outflow, \citet{1999A&A...351.1103R}) whereas others 
(e.g. VLA 0548) 
show wind-driven signatures \citep{2000ApJ...542..925L,2001ApJ...557..429L}.
However, when massive star formation is considered, the observations cannot
be matched by any of these two models 
\citep{1997ApJ...479L..59C,2000upse.conf...41C}. 
The formation of massive stars gives rise to molecular outflows 
showing different properties than
the ones recorded in low and intermediate mass star formation: they are
very massive ($\sim$~tens of solar masses and, in some cases,
more massive than the 
central stars that presumably drive them), poorly collimated and
also faster than in the case of low mass star formation. 
 According to \citet{1997ApJ...479L..59C,2000upse.conf...41C},
jets cannot entrain more than a few solar masses 
and certainly not the tens of solar masses observed in massive outflows.  
Furthermore,
the wide opening angles observed are hardly explained by the 
entrainment from a collimated jet. On the other hand, if entrainment
by a wide-angle wind allows a large opening angle for the 
outflow, it cannot match the velocities reached by massive
outflows. 

In this paper, we present a general radial self-similar model for the 
flows surrounding young forming stars, applicable in both low and 
high mass star formation. The paper is laid out as follows: 
in Sect.~\ref{theory}, we describe the model and its equations and show 
corresponding hydrodynamical solutions in Sect.~\ref{HD}; 
some magnetized solutions are then presented, together with a study of
the influence of changes in the opacity, in Sect.~\ref{MHD}; 
we give the main properties of the model in Sect.~\ref{Gprop} and
discuss them in Sect.~\ref{discussion} before 
concluding.  


\section{Theoretical basis of Transit Models \label{theory}}
 The present models are based on ideal MHD
with the assumption of radial self-similarity and a 
simplified treatment of radiation \citep{FH96, lery03}. 
 
Spherical coordinates are used: the system
is centered on the protostar, the $x-y$ plane corresponds to the
equatorial plane and the poloidal angle $\theta$ is taken relative
to the rotational axis of the central object.

\subsection{Fluid equations}
We describe the gas with the usual 
macroscopic quantities: density $\rho$,
velocity ${\rm \bf v}$, pressure $p$ and temperature $T$.
The size of the system is such that the conditions for ideal MHD
are fulfilled.  We further assume steady-state ($\partial/\partial t =0$), 
axial symmetry around the rotational axis ($\partial/\partial \phi =0$).
The standard ideal MHD equations then reduce to:

\begin{equation}
\nabla\cdot(\rho \mathbf{v})=0
\label{continuity}
\end{equation}

\begin{equation}
\frac{1}{\rho}\nabla p +\nabla \Phi+\left(\mathbf{v \cdot \nabla}\right)
\mathbf{v}=
\frac{1}{4\pi\rho}\nabla \times \mathbf{B} \times \mathbf{B}
\label{momentum}
\end{equation}

\begin{equation}
\nabla\cdot \mathbf{B}=0 \; .
\label{magneticflux}
\end{equation}

The magnetic Reynolds number being much greater than unity,
the diffusion term in the induction 
equation can be neglected, and the magnetic field lines are
frozen in the plasma.
The induction equation becomes
\begin{equation}
{\nabla\times(\mathbf{v} \times \mathbf{B})=0}
\label{induction}
\end{equation} 
which directly implies
\begin{equation}
\mathbf{v} \times \mathbf{B}=\nabla \Psi
\label{electricpot}
\end{equation}
where $\Psi$ is the electric potential (${\mathbf{E}=-\nabla \Psi}$).
A non-zero toroidal electric field could convect 
the magnetic poloidal field lines into a sink on the axis of the system.
To ensure a zero toroidal electric field, the
only requirement from Eq.~(\ref{electricpot}) 
is $\mathbf{v}_p\propto\mathbf{B}_p$ 
\citep{1980ApJ...241..534C, henriksen96}, where
the subscript $p$ denotes the poloidal component of the field 
(${\mathbf{v}=\mathbf{v}_p+\mathbf{v}_{\phi}}$ 
and $\mathbf{B}=\mathbf{B}_p+\mathbf{B}_{\phi}$).

We also consider the equation of state for ideal gas, i.e. $p=nkT$. 

Note also that the purely hydrodynamical problem will be studied 
by using Eq.(\ref{continuity}) and setting the RHS in Eq.(\ref{momentum}) 
to zero.

\subsection{Self-similar treatment}

\subsubsection{Self-similarity - general statements}

 A phenomenon is called self-similar if the spatial (or temporal)
distributions of its properties at various different times (or locations)
can be obtained from one another by a similarity transformation.
The investigation of the full phenomenon can then be reduced 
to the study of the properties of the system for only a specific
time (or location). If the origin of time can be chosen arbitrarily, 
the scales of length and mass are also arbitrary, and the system is 
`scale-free'. 
The system of partial 
differential equations (PDE) that describe the problem is transformed to a set
of ordinary 
differential equations (ODE), which drastically simplifies the investigation.

Self-similar models turn out not only to describe the behavior of
physical systems under some special conditions,
but also describe the intermediate-asymptotic
behavior of solutions to wider classes of problems in the range where
these solutions no longer depend on the details of the initial
and/or boundary conditions, yet the system is still far
from being in an ultimate equilibrium state \citep{1972AnRFM...4..285B}.
For an extensive review of self-similarity, the reader is referred to 
\citet{self-sim}. In present case of star formation, 
we express each variable using the following self-similar form
\begin{equation}
L(r,\theta)=L_0 \times \left(\frac{r}{r_0}\right)^{\alpha} \times l(\theta).
\label{selfsim}
\end{equation}
The constant $L_0$ has the dimensions of the variable and is
directly dependent on the parameters of the problem 
such as the mass or the luminosity of the central object.
The free parameters are the self-similar indices $\alpha$ and the fiducial
scale $r_0$.  

\subsubsection{Self-similar fluid quantities}
 Several observations of the protostellar environment favor 
our self-similar approach since they show that radial density profiles 
in regions of isolated star formation are
well represented by a power-law $\rho \propto r^{-p}$ with $p \sim 0.5-2$ 
(e.g. L1527 
has been found to have $p$ in the range of 1.5-2 \citet{1991ApJ...382..555L}). 
Indeed, in some cases, density profiles are reproduced by most 
hydrostatically supported ($p=2$)
or free infalling ($p=1.5$) cloud core, e.g. \citet{1977ApJ...214..488S}. 
In some other cases, the slope has
been found to be much shallower than the ones predicted 
by the standard models. 
Density profiles as shallow
as 0.5-0.9 can be interpreted as the presence of a magnetic   
\citep{1993ApJ...406L..71B} or rotational support\citep{1998MNRAS.299..789C} 
of the core.

We now apply our assumptions to the ideal MHD equations. 
Dimensional analysis allows us to
determine the index of each variable as a function of a single
free self-similar parameter, $\alpha$.
The systems then reads
\begin{equation}
\rho(r,\theta)=\frac{M}{r_0^3}\left(\frac{r}{r_0}\right)^{2\alpha-\frac{1}{2}}
\mu(\theta)
\label{self-density}
\end{equation}
\begin{equation}
p(r,\theta)=\frac{GM^2}{r_0^4}\left(\frac{r}{r_0}\right)^{2\alpha-\frac{3}{2}}
P(\theta)
\label{self-pressure}
\end{equation}
\begin{equation}
T(r,\theta)=\frac{\bar{\mu} m_H}{k_B}
\frac{GM}{r_0}\left(\frac{r}{r_0}\right)^{-1} \Theta(\theta)
\label{self-temperature}
\end{equation}
\begin{equation}
\mathbf{v}(r,\theta)=\left(\frac{G M}{r_0}\right)^{\frac{1}{2}}
\left(\frac{r}{r_0}\right)^{-\frac{1}{2}} \mathbf{u}(\theta)
\label{self-velocity}
\end{equation}
\begin{equation}
B_{r,\theta,\phi}(r,\theta)
=\left(\frac{G M^2}{r_0^4}\right)^{\frac{1}{2}}
\left(\frac{r}{r_0}\right)^{\alpha-\frac{3}{4}}
\frac{u_{r,\theta,\phi}(\theta)}
{y_{p,p,\phi}(\theta)} .
\label{self-magnetic}
\end{equation}

In these expressions, $G$ is the gravitational constant, $k_B$ the
Boltzmann constant, $m_H$ the mass of the proton, $\bar{\mu}$ the
mean molecular weight (we use $\bar{\mu}=2$), and $M$ the mass of
the central object. The fiducial scale $r_0$ is a radius of reference. 
The self-similar assumption is valid only above this radius  
(see Sect.~\ref{r0}). 
Note also that in Eq.~(\ref{self-magnetic}), the 
subscript $p$ stands for ``poloidal'.
The self-similar index $\alpha$ is a \emph{free} parameter that lies between 
$-1/2<\alpha\leq 1/4$. In particular,
$\alpha=-1/2$ yields to pure radial accretion. This issue is discussed
in \citet{FH96} and the behavior of the solutions with $\alpha$ can be
found in \citet{Lery99}. Note that the hydrostatic ($p=-2\alpha+1/2=2$) and 
the free infalling ($p=1.5$) cases cannot be treated by the present model.
 
For the remainder of this paper, we choose
$\alpha=-0.2$ in most cases. 
From Eq.(\ref{self-density}), we infer
$\rho \propto r^{-0.9}$, which corresponds to a shallow density profiles
where rotation and magnetic field can drive substantial effects.

In a nearly study on self-similar transit models \citep{FH96}, 
the same proportionality relationship between each component of the velocity
and magnetic field was taken.
All the components of the electric field Eq.(\ref{electricpot}) 
were then equal to zero.
Such a configuration has the virtue of great simplicity, but
does not allow for the existence of a Poynting flux. They obtained fast 
axial outflows but the luminosities required to drive them were as
high as $10^5-10^6 \; L_\odot$.
In a subsequent work \citep{Lery99}, this constraint has been relaxed. 
Only the necessary proportionality condition
on the poloidal components of the velocity and magnetic field 
(${\bf v}_p \propto {\bf B}_p$) 
was satisfied and the importance of
the Poynting flux taken into account. The corresponding solutions
were found to be faster and required a comparatively 
smaller source luminosity to be driven. 
In this work, 
we use the formalism and conditions from \citet{Lery99} 
based on Poynting flux driving mechanism 
($y_p(\theta) \neq y_\phi(\theta)$)
that breaks the collinearity between ${\bf B}$ and ${\bf v}$.

\subsubsection{Self-similar radiation treatment \label{rad}}
 In the early stages of star formation, 
most of the luminosity comes from the accretion shock created 
by the infalling material.  Later on,
the accretion rate reduces and 
the radiation is dominated by the protostar luminosity.
Here, we use a simplified description for radiation \citep{FH96} 
that allows us to use an analytical self-similar approach.
\paragraph{Radiative diffusion:}
The steady-state energy conservation equation, that  includes 
the mechanical and radiative energy fluxes, simply reads
\begin{equation}
\nabla\cdot\left[\rho {\bf v}\left(h+h_{rad}+\frac{v^2}{2}+\Phi_{grav}\right)
+{\bf F}_{rad} \right]=0
\label{nrjconserv}
\end{equation}
where:
\begin{itemize}
\item $h$ is the specific enthalpy: $h=u+p/\rho$, with $u$ 
the internal specific energy 
and $p$ and $\rho$ respectively the pressure and density.
\item $h_{rad}$ is the specific radiative enthalpy. For an isotropic radiation 
field, the radiative pressure is $p_{rad}=u_{rad}/3$, with $u_{rad}$ the 
radiative energy density. Then, $h_{rad}$ is defined as 
$h_{rad}=(u_{rad}+p_{rad})/\rho=4p_{rad}/\rho$
\item $\Phi_{grav}$ is the gravitational potential created by the 
central protostar.
\item ${\bf F}_{rad}$ is the radiative flux.
\end{itemize}
For the radiation independently, the radiative energy conservation reads 
\citep{1982JCoPh..46...97M}
\begin{displaymath}
\nabla \cdot \left({\bf F}_{rad} +\rho {\bf v} h_{rad}\right)=
-\frac{\kappa\rho}{c} {\bf v}\cdot {\bf F}_{rad}
\end{displaymath} 
where $c$ is the speed of light and $\kappa$ the opacity.
It is re-written at the zero-\emph{th} order in (v/c)
\begin{equation}
\nabla \cdot {\bf F}_{rad}=0 \;.
\label{radiation}
\end{equation}
This equation decouples the mechanical energy fluxes from the radiative
ones in Eq.(\ref{nrjconserv}).
Furthermore, when taking the zero-\emph{th} order in (v/c) in 
Eq.(\ref{nrjconserv}), the $h_{rad}$ term disappears.
This means that the radiative pressure is neglected,
i.e. the photons transfer no momentum to the gas.  

The previous statements are valid for any particular form of
the radiative flux ${\bf F}_{rad}$. In this work, we use the
radiative diffusion approximation 
\begin{equation}
\mathbf{F}_{rad}=-\frac{c}{\kappa \rho} \nabla p_{rad}
\label{frad}
\end{equation}
where the radiative flux 
is directly linked to the energy density (then to the radiative pressure). This
assumes that the mean free path of a photon is very short compared to 
the characteristic length of the system. 
We use the black-body radiative pressure for $p_{rad}$ which, 
at the temperature $T$ is given by
\begin{equation}
p_{rad}=\frac{4}{3}\frac{\sigma}{c}T^4
\label{prad}
\end{equation}
with $\sigma$ the Stefan-Boltzmann constant.

\begin{figure}[t]
\begin{center}
\includegraphics[width=7cm]{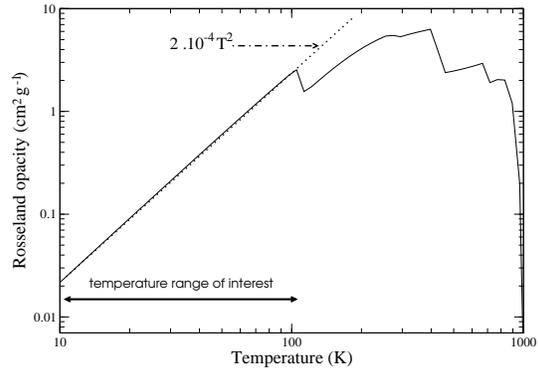}
\end{center}
\caption{Opacity as a function of temperature computed with
the code from \citet{2003A&A...410..611S} for a gas density of 
$10^{-18}$g~cm$^{-3}$.\label{opacity}}
\end{figure}

\paragraph{Opacity:}
For the sake of simplicity, we
choose an opacity defined by Kramer's law, 
\begin{equation}
\kappa=\kappa_0 \left(\frac{\rho}{1 \rm{~g~cm}^{-3}}\right)^a
\left(\frac{T}{1 \rm{~K}}\right)^b \; ,
\end{equation}
where the values of the exponents 
$a$ and $b$ are defined according to  
the type of coolant considered. These parameters will be discussed
in greater detail in section \ref{kramer}. For example, when the
cooling is dust-dominated, $a=0$ and $b=2$ (e.g. \citet{1985Icar...64..471P}).
The constant $\kappa_0$ is determined using a numerical code developed
by \citet{2003A&A...410..611S} which calculates 
opacities over a wide range of gas densities and temperatures, given 
a solar-type metallicity. This code includes different shapes
and compositions for the dust particles. In this work, we
use the ``homogeneous compact
spherical dust'' configuration to compute the opacity.
Moreover, we focus on molecular outflows which temperature range (10-100~K) 
is in the dust-dominated regime of opacity. In that case, the dust being such
an efficient coolant, the opacity
is independent of the density of the gas, i.e. $a=0$. The opacities, obtained 
for a gas density of $10^{-18}$g~cm$^{-3}$,
are plotted in Fig.~\ref{opacity} and can be fitted by a power-law. 
We found the Kramer's coefficient of the temperature to be $b=2$ and 
$\kappa_0 \sim 2\times 10^{-4}$~cm$^2$~g$^{-1}$ for the dust dominated regime.

Following Eq.(\ref{selfsim}), the self-similar radiative flux reads
\begin{equation}
\mathbf{F}_{rad}(r,\theta)=\left(\frac{G M}{r_0}\right)^{\frac{3}{2}}
\frac{M}{r_0^3}
\left(\frac{r}{r_0}\right)^{\alpha_f -2} \mathbf{f}(\theta) \; .
\label{self-frad}
\end{equation}
Using the self-similar expression of the temperature $T$ in Eq.~(\ref{prad}) 
and combining Eq.~(\ref{frad}) and (\ref{prad}), one finds 
$\alpha_f$ to be $\alpha_f=b-3+2(a+1)(1/4-\alpha)$. 

Furthermore, the inclusion of radiative diffusion in the model
constrains the expression of the fiducial scale $r_0$. Indeed, the RHS
and LHS multiplicative constants of the self-similar expansion of 
Eq.~(\ref{frad}) have to be equal: this leads to the following expression 
of $r_0$,
\begin{equation}
r_0=\left(\frac{3\kappa_0}{4a_sc}M^{2+a}
\left(GM\right)^{b-\frac{5}{2}}
\left(\frac{\bar{\mu}m_H}{k_B}\right)^{b-4}\right)^{\frac{2}{5+6a+2b}}
\label{fiducial}
\end{equation}
where $a_s=4\sigma/c$ with $\sigma$ the Stefan-Boltzmann constant.

It is worthwhile to note that when diffusive radiation is 
discarded\footnote{i.e., 
there is no $\theta$-component in the radiative flux, but only a 
radial one, cf. the ``virial isothermal'' case in 
\citet{FH96}.}, the fiducial scale is not explicitly defined. In that case, 
the fiducial radius remains a free parameter and 
a characteristic length has to be chosen on observational or physical 
criteria (see \citet{Lery99} and Sect.~\ref{r0} 
for the discussion on this issue).

\subsection{Methods}

\subsubsection{Integration\label{integration}}
Assuming self-similarity, the PDE system of 
HD/MHD equations is transformed into an ODE system
where all variables depend only on the poloidal angle
$\theta$,
\begin{equation}
\frac{{\rm d}l_i(\theta)}{{\rm d}\theta}=A_{ij}\; l_j(\theta)\; .
\end{equation}
In the MHD (HD) case, eight
(six) coupled equations constitute the system.
We consider the problem as an initial value problem (IVP) and
use the values of the variables ($u_r,u_\theta,
u_\phi, \mu, y_p, y_\phi, \Theta$ and $f_\theta$) as input parameters 
at a given angle. Note that $ y_p$ and $y_\phi$
are not present in the HD case.
In practice, we start the integration close to the rotational axis,  
typically at $\theta_0=10^{-2}$ rad.
We provide the initial values near the axis, i.e. a positive
$u_r$ for the outflow, $u_{\theta}$ negative for the 
circulation pattern and $u_{\phi}$ chosen to be positive.
We look for solutions covering $\theta=\theta_0$ to 
$\theta=\pi/2-\epsilon$ as the polar axis and the equatorial
plane are excluded by the self-similar treatment \citep{FH96}. 
%

\subsubsection{Dimensions and the 
fiducial scale r$_0$\label{r0}}
For each solution, 
dimensional quantities can be calculated using Eq.(\ref{self-density})
to (\ref{self-magnetic}) and Eq.(\ref{self-frad}).
From these equations, for any physical quantity, only
the mass of the central object $M$ and the fiducial scale $r_0$
are  required to compute the dimensional quantities 
at a given radius $R$.
Without radiative diffusion, $r_0$ remains a free parameter 
(see \citet{Lery99} for more details).

\begin{figure}[ht]
\begin{center}
\includegraphics[width=8cm]{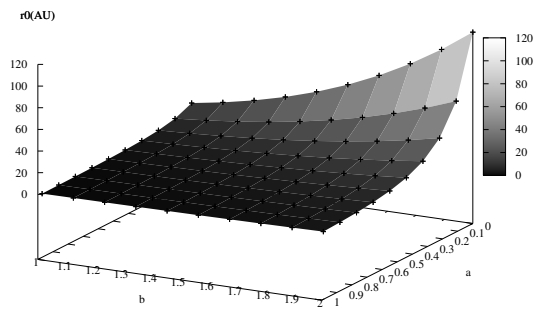}
\caption{Variation of $r_0$ with the Kramer's opacity parameters.
The values are obtained for a one solar mass protostar.
\label{plot_r0}}
\end{center}
\end{figure}

In Fig.~\ref{plot_r0}, the fiducial scale (from Eq.(\ref{fiducial}))
is plotted as a function 
of the Kramer's coefficients $a$ and $b$ for a one solar mass protostar
and $\kappa_0=2\times10^{-4}$cm$^2$~g$^{-1}$.

The variation of $r_0$ with the Kramer's parameters is quite stiff and 
$r_0$ quickly drops from $\sim$ 100~AU to 0 as $a$ increases 
and/or $b$ decreases from the dust-dominated case.
Note that $r_0$ is a lower
limiting radius for the validity of self-similarity.

For physical reasons, $r_0$ should also be limited to the region
where the hypotheses of the model still apply.
For example, close to the protostar, the radiation
pressure could have non-negligible dynamical effects preventing the model
to be applicable there. 

\subsubsection{Selection criteria of the solutions}

Firstly, we select solutions showing outflowing motions near
the axis region and infall near the equatorial plane. This means 
that radial velocity $u_r$ must change sign once for an angle 
$\theta_{\rm open}$ which is the opening angle of the outflow: for 
$\theta>\theta_{\rm open}$, $u_r$ is negative and trace the infall;
for $\theta<\theta_{\rm open}$, $u_r$ is positive and the gas 
is flowing outward.

Secondly, we use the self-similar expressions of the physical
quantities to go from the dimensionless variables to the
dimensioned ones: we define the mass of the
central object (we typically choose 1~$M_\odot$), an 
observational radius ($\sim 5 \times 10^3$~AU) and a type
of coolant (values of $a$ and $b$ in the opacity).

Finally, we discard unphysical solutions, i.e. solutions that
do not fit in the range of observations.

\section{Hydrodynamical solutions\label{HD}}

Pure hydrodynamical (HD) solutions can apply in situations where 
magnetic fields are either very weak (e.g., primordial universe)
or not dynamically dominant for the movement of the gas (early
stages of the collapse).
Besides, studying a purely HD problem allows us
to distinguish the effects of the radiation and density gradients
from those of the magnetic field (see Sect.~\ref{MHD}).

We find HD solutions with both infall and outflows and 
they all show common properties: i) the outflow is very narrow
and ii) always quite slow ($<10$~km~s$^{-1}$ for a one solar
mass central object) and iii) infall is occurring in the rest of the
domain in a quasi-spherical way.

\begin{figure*}[t]
\begin{center}
   \begin{minipage}[c]{.48\linewidth}
     \includegraphics[clip=, bb=36 145 532 595,width=8cm]{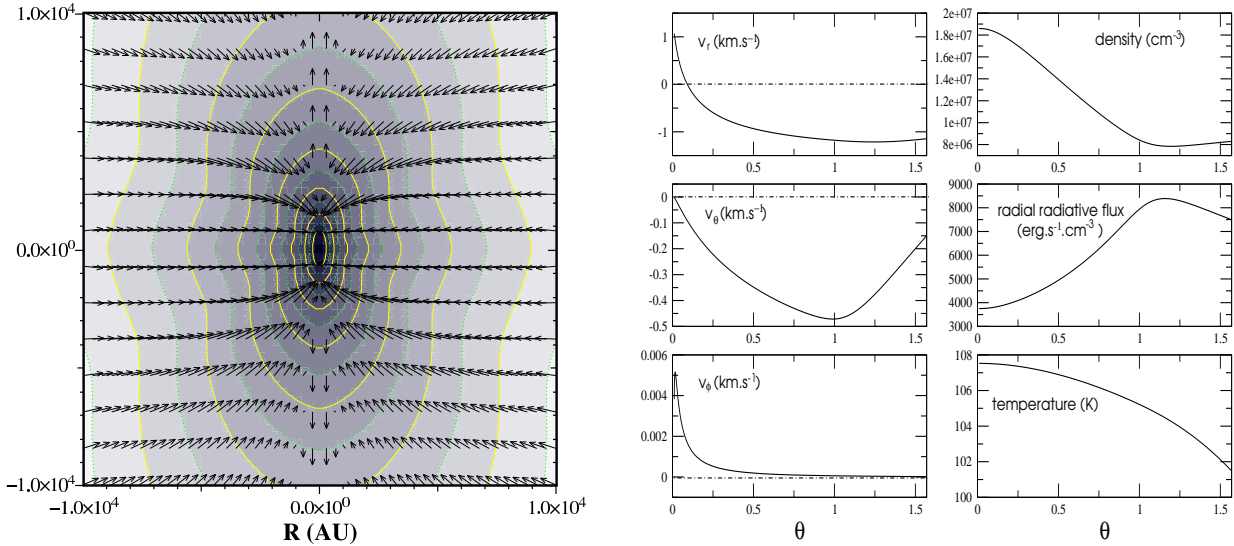}
   \end{minipage} \hfill
   \begin{minipage}[c]{.48\linewidth}
      \includegraphics[height=7.1cm,width=7.9cm]{f3b.eps}
\end{minipage}
\caption{Left panel: a typical solution for the HD case represented
in the poloidal plane $(r,\theta)$. The iso-density contours and 
velocity field are plotted. 
Right panel: corresponding values of the velocity, density, 
temperature and radial radiative flux at a distance of 5000~AU
from the central protostar, with respect to the poloidal angle $\theta$.
For this particular solution, the input values of the dimensionless
quantities are: the velocities $u_r=2.54$, $u_\theta=-4.46\times10^{-3}$,
$u_\phi=9.05\times10^{-3} $, the density $\mu=3.62\times10^{-3}$, 
the temperature $\Theta=1.81$ 
and the $\theta-$component of the radiative flux $f_\theta=7.92\times10^{-1}$.
\label{solhydro}}
\end{center}
\end{figure*}
 
For illustration, a typical hydrodynamical solution is plotted in
Fig.~\ref{solhydro}: the left panel corresponds to the density contours
and velocity field of the solution in the poloidal plane, and the right
panel to different quantities plotted (at a fixed distance of 5000~AU)
as a function of the poloidal angle $\theta$. The central object
was assumed to be a one solar mass protostar. The Kramer's opacity 
parameters are those of a dust-dominated opacity (namely, $a=0$ and $b=2$). 

The infall-outflowing pattern is traced by the radial velocity: a negative 
radial velocity corresponds to an infall motion whereas a positive
one is the signature of an outflow. Note also that $v_\theta$
keeps a constant negative sign, implying that all the material is being
redirected outward in this particular solution. 
This will not be the case for solution
with ``net infall'' (see Sect.~\ref{typicalMHD}). The $\theta-$velocity
characterizes the movement of the gas towards the rotational axis.
Here, the maximum reached by $v_\theta$ is $\sim$~0.45~km~s$^{-1}$: this is a
small value that implies a weak ``collimation'' of the flow.
For this solution, 
the outflow has a velocity of approximatively $\sim$~1~km~s$^{-1}$ 
and a $\sim$~5$^\circ$ opening angle. The system is also almost
non-rotating as $v_\phi$ is close to zero over the entire domain. 
The values of the number density (upper-left panel) are in the
order of $10^6$~cm$^{-3}$ at 5000~AU and 
shows a increase towards the axis of rotation. However,
the density at the axis is only twice as large as the one near the equator,
emphasizing a quasi-spherical infall. 
At a given radius, the temperature (lower right panel) is also almost
constant ($\sim$~100~K at 5000~AU) which characterizes an isothermal infall.

Qualitatively, this solution presents some common features with observations.
First we have an outflow. We also find that the fastest material was also the most collimated
\citep{1996ARA&A..34..111B,2000prpl.conf..867R}.
This is precisely the case here where the positive part of the radial velocity 
(outflow) that decreases when the angle from the axis increases.
However, physical quantities (velocity, temperature and density) do not match
with observations. 
In particular, the density is an order of magnitude too high to fall in 
the observed density range
($10^4-10^5$~cm$^{-3}$), and also the velocity is too small. 
 
In conclusion, we have shown that our models can produce 
outflows that can be launched from thermodynamical effects only. 
We have a simple heated quadrupolar model for infall and outflow.
The qualitative characteristics of the hydrodynamical solutions
are similar to the observed protostellar outflows but are 
quantitatively too dense, too slow and too narrow for typical 
solar mass objects.

\section{MHD solutions\label{MHD}}

Typical molecular outflows of low mass protostars 
have observed velocities 
$\sim$~20~km~s$^{-1}$ and have a large range of initial opening angles:
from $\lesssim$~30$^\circ$ for class 0 to $\gtrsim$~90$^\circ$ for class 1
objects \citep{1996ARA&A..34..111B,1999osps.conf..227B}. 
Their typical densities lie 
around $10^{4-5}$~cm$^{-3}$. 
Our hydrodynamical model cannot power such outflows 
and we now include the magnetic field in the model.

There are two main issues that we wish to address with
our magnetized model:
\begin{itemize}
\item Class 0 low mass protostars: they show powerful, highly collimated
molecular outflows despite their very young age.
\item The formation of massive stars: they present very massive outflows
(mass loss rates $\sim 5 \times10^{-3}$~$M_\odot$~yr$^{-1}$) that can 
be more massive than the central object itself and 
hardly be explained by the conventional jet- or wind-driven models 
\citep{2000upse.conf...41C}.
\end{itemize}

\subsection{Dust case -- a typical solution \label{dust}}

\begin{figure*}[t]
\begin{center}
   \begin{minipage}[c]{.48\linewidth}
      \includegraphics[clip=, bb=36 145 532 595,width=8cm]{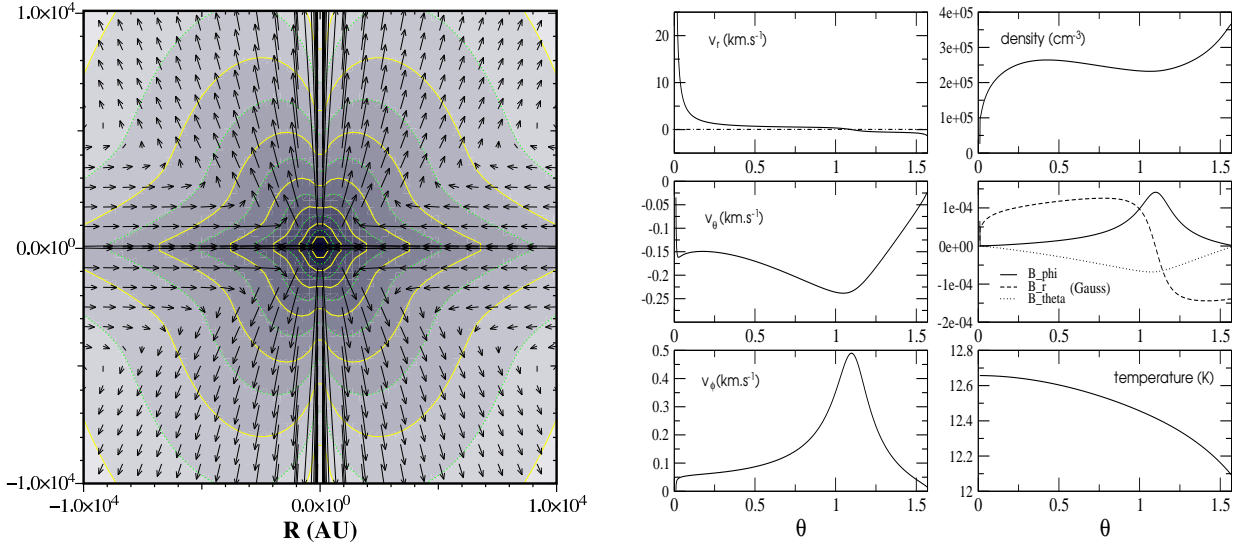}
   \end{minipage} \hfill
   \begin{minipage}[c]{.48\linewidth}
      \includegraphics[height=7.1cm,width=7.9cm]{f4b.eps}
\end{minipage}
\caption{Same as Fig.~\ref{solhydro} but for a MHD solution 
with pure transit.
For this particular solution, the initial values of the dimensionless
variables are: $u_r=1.45\times 10^3$, $u_\theta=-8.12\times 10^{-2}$, 
$u_\phi= 7.49\times 10^{-3}$, $y_p= 5.12 \times 10^5$, 
$y_\phi=5.81\times 10^{-1}$, $\mu=5.93 \times 10^{-6}$, 
$\Theta=2.13\times 10^{-1}$
and $f_\theta=1.353$. The value of $\alpha$ is -0.22 and the 
opacity parameters correspond to the dust case $a=0$ and $b=2$.
\label{solcirc}}
\end{center}
\end{figure*}

A typical MHD solution with pure transit is shown in Fig.~\ref{solcirc}.
The left panel corresponds to the density contours and velocity field
in the poloidal plane and the right panel to different quantities
plotted at a fixed radius of 5000~AU as a function of the angle $\theta$.

As previously, the radial velocity traces the infalling/outflowing motion
of the gas. For this solution, the opening angle is
$\theta_{\rm open}=1.1 \; {\rm rad} \approx 60^\circ$, where 
$v_r$ changes sign. However, between 1.1 and 0.3, the radial velocity
is positive but very small and it is only beyond 0.3~rad~$\approx 17^\circ$
that a non negligible outflowing motion is occurring.

The outflow velocity is typically 10-20~km~s$^{-1}$ for
the most collimated part and the number density varies from $10^5$ to 
$4\times 10^5$~cm$^{-3}$ between $0<\theta<\pi/2$. These values fit in the
range of typical values of molecular outflows \citep{1996ARA&A..34..111B}.
The turning angle is the preferential angle for the toroidal magnetic
field (middle right panel), responsible of the collimation of the flow
\emph{via} the $j_z \times B_\phi$ component of the Lorentz force. In
consequence, $v_\theta$, which traces the movement of the gas towards
the axis, is also maximised at the turning point. Naturally, 
this correlation was not found in the HD case where the 
opening angle was very small but $v_\theta$ was maximum around 55~$^\circ$. 
Remembering that $B_\phi \propto v_\phi$ in the model, $v_\phi$
behaves identically to the toroidal field.

The density (upper right panel) presents a strong poloidal gradient 
($\partial_\theta$) near the 
rotational axis, followed by an almost constant behavior for
intermediate values of $\theta$ and finally increases again when
approaching the equatorial region. The strong gradient near
the axis contributes to the acceleration of the outflowing
material in this region. This behaviour is not present in the 
HD solutions.
  The density structure of the present 
MHD solution appears oblate with respect to the rotational axis 
whereas it looks prolate for the hydrodynamical model. 
Such a geometrical difference may be observationally distinguishable 
at high resolution. Indeed \citet{motte01} have investigated 
the morphologies of Class 0 and 1 protostellar envelopes and have found that 
several sources (e.g. L1527, B335) have elliptical (or even more complex) 
density structure. They invoke the presence of bipolar outflows to possibly 
be the reason of these asymmetries. This could be the case in our model. 

\begin{figure*}[ht]
\begin{center}
   \begin{minipage}[c]{.48\linewidth}
 	\includegraphics[clip=, bb=36 145 532 595,width=8cm]{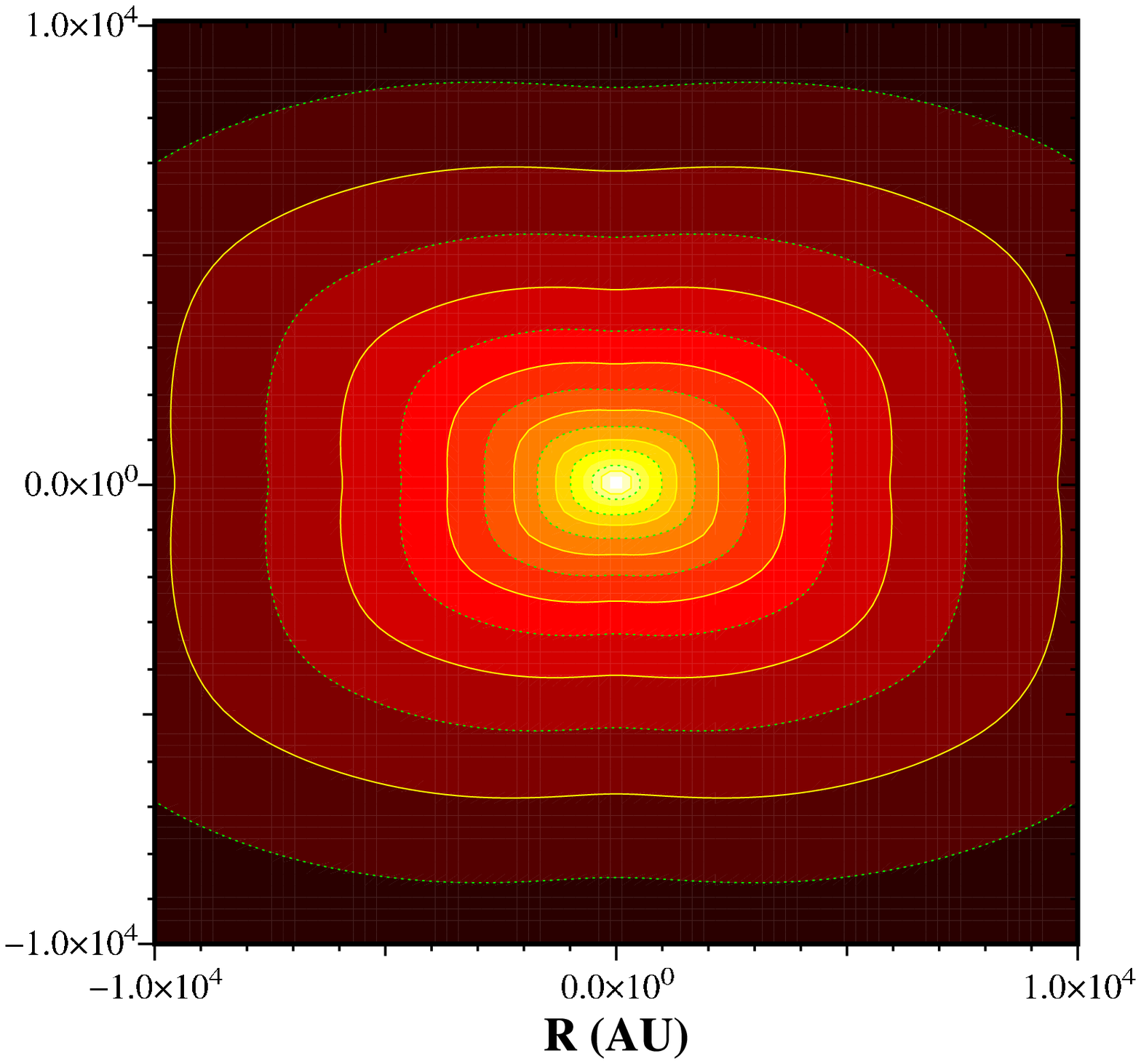}
   \end{minipage} \hfill
   \begin{minipage}[c]{.48\linewidth}
        \includegraphics[clip=, bb=36 145 532 595,width=8cm]{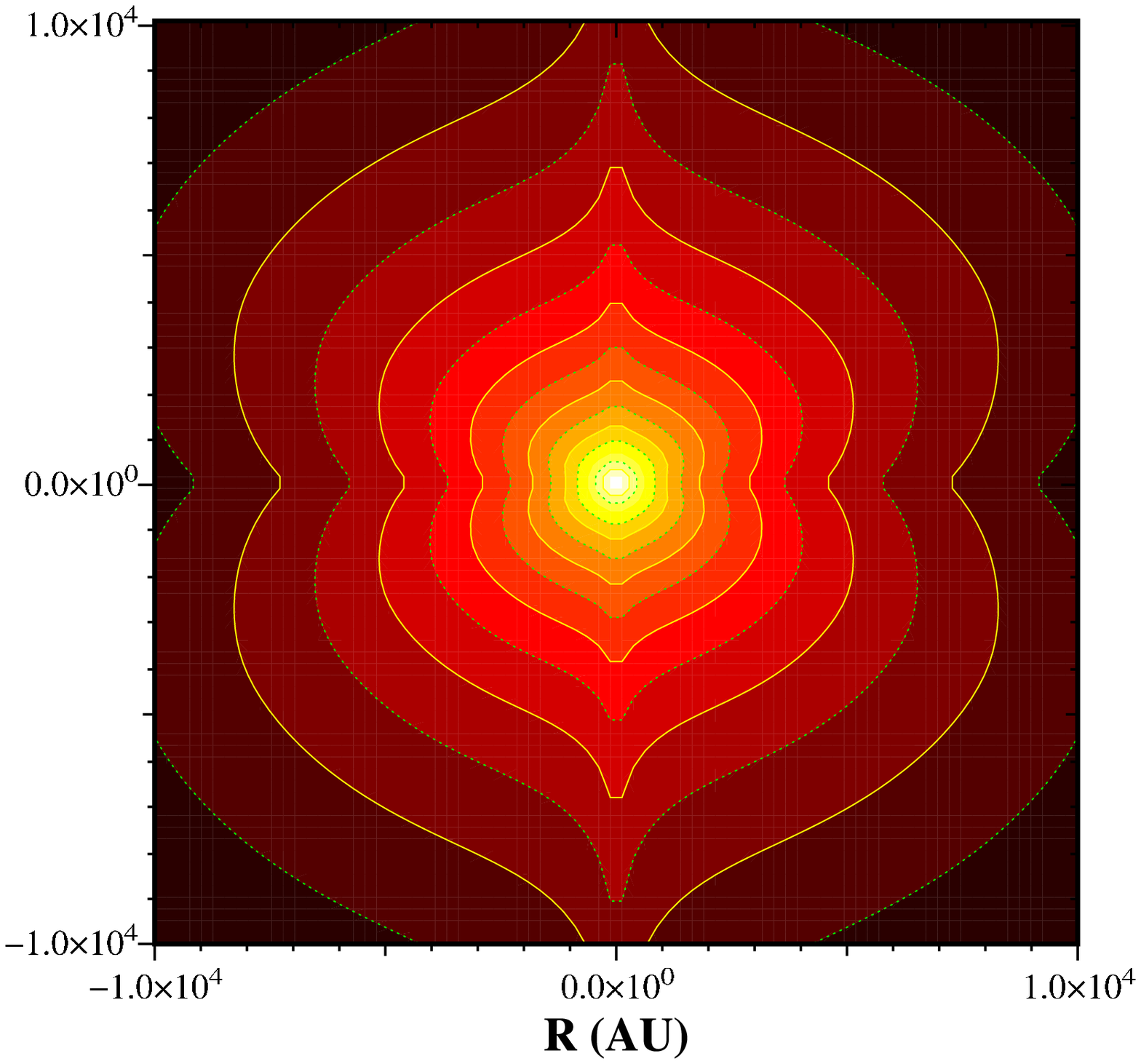}
   \end{minipage}
\caption{The radiative flux contours of typical transit solutions in the 
HD (left) and MHD (right) case between 
0 and $10^4$~AU. At 5000~AU, the typical values
are respectively $6\times 10^3$ and $3\times 10^3$~erg~s$^{-1}$~cm$^{-2}$. 
\label{Frad_B_ssB}}
\end{center}
\end{figure*}

As in the hydrodynamical case, the temperature is almost constant
over the domain. However, the radiative flux plotted in the poloidal plane
in Fig.~\ref{Frad_B_ssB}, right panel, shows a strong anisotropy and is 
the highest in the axial zone and also contributes to the acceleration
of the outflowing gas. On the other hand, the flux appears smaller in
the equatorial region than elsewhere, emphasizing that the equator
is a privileged region for the accretion of  matter onto the central protostar.
That is not the case for the HD model (Fig.~\ref{Frad_B_ssB}, left panel) where
the radiation is almost spherical.

The typical values for two models, hydrodynamical and magnetized, are
gathered in Tab~\ref{compare_B_nonB} which summerizes the previous points.

\begin{table*}[ht]
\begin{center}
\begin{tabular}{r c c c c c c}
\hline\hline
  & $v_r^{\rm out}$ (km.s$^{-1}$)
  &$v_\theta^{\rm max}$ (km.s$^{-1}$)
  & $v_\phi^{\rm max}$ (km.s$^{-1}$)
  & $\rho_{\rm out}$ (cm$^{-3}$)&$\rho^{\rm axis}/\rho^{\rm eq}$
  &$F_{\rm rad}^{\rm axis}/F_{\rm rad}^{\rm eq}$\\
\hline
with {\bf B} field
& $\sim 10$ & 0.25  & 0.5 & $2 \times 10^{5}$ & 0.25  &4.3 \\
no {\bf B} field & $\sim 1$ & 0.45 & $5 \times 10^{-3}$ & $1 \times 10^{7}$ 
& 2 & 0.5 \\
\hline
\end{tabular}
\caption{Comparison between typical values of the HD and MHD
solutions at a distance of 5000~AU from the central protostar.  
\label{compare_B_nonB}}
\end{center}
\end{table*}

Knowing the velocity and the density for every angle $\theta$ at a fixed
radius R, it is possible to calculate mass rates. In particular, the infall
mass rate is given by 
\begin{equation}
\dot{M}_{\rm in}=2 \times 2\pi R^2 \int_{\theta_{\rm open}}^{\pi/2}
v_r(R,\theta) \rho(R,\theta) \sin \theta {\rm d}\theta
\label{infall_rate}
\end{equation}
where $v_r(R,\theta)$ and $\rho(R,\theta)$ takes the form of 
Eq.(\ref{self-velocity}) and Eq.(\ref{self-density}).
The first factor 2 is present to take into the contribution to the infall
from the other side of the equatorial plane.

\begin{figure}[th]
\begin{center}
\includegraphics[bb=50 50  730 580, width=8cm]{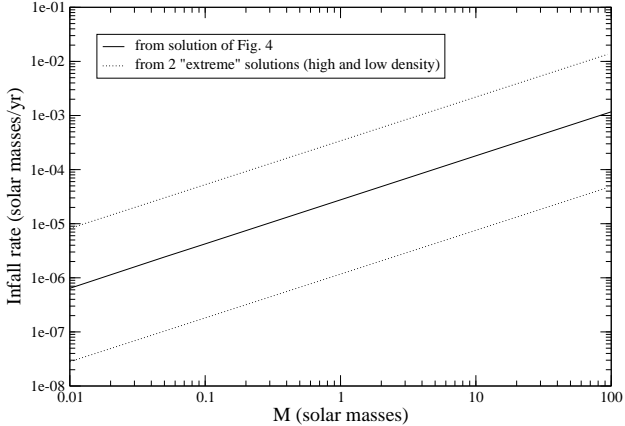}
\caption{Influence of the central mass on the infall rate for a MHD transit
solution. The solid line corresponds to the specific solution 
presented previously. The dotted lines correspond 
to two ``extreme'' solutions of high and low density.\label{influence_M}}
\end{center}
\end{figure}

We compute the infall rates for different central masses and plot the result
in Fig.~\ref{influence_M}. Three solutions are treated: the one detailed
previously in Fig.~\ref{solcirc} (solid line) 
and two extreme cases (a low and a high density solutions,
in dotted lines). The low and high density solutions are extreme in 
the sense that they are at the limit of the observational range
but not a limit of the model (which can also produce denser or lighter
solutions). 
Then, for a given central mass, our model predict
infall rates located between the two dotted lines.

First of all, the infall rate increases with the mass of the protostar
and we get a typical value of 
$\dot{M}_{\rm in} \approx 3 \times 10^{-6}$~$M_\odot$~yr$^{-1}$ 
for a 0.1~$M_\odot$
central object, which is in agreement with rates inferred 
from observations of low mass YSO \citep{1996A&A...311..858B}. 
It is important to mention here that 
the increase of the central mass does not 
correspond to the temporal evolution of the mass of the protostar:
we are studying a self-similar steady-state problem where the central mass
is a dimensional parameter. Observationally, the higher the central mass is,
the later in the pre-stellar evolution the protostar is, and as a consequence,
the lower its accretion rate is. Here we are just studying the influence
of the central mass on a given model and the increase of the infall
rate is a consequence of the increase of the gravitational force
from the source. 

The infall rate scales with $M$ following a power-law. For this example,
we have $\dot{M}_{\rm in}\propto M^{0.81}$. This behavior comes
from the self-similar form we use for the variables and, injecting 
Eq.(\ref{self-density}) and Eq.(\ref{self-velocity}) 
into Eq.(\ref{infall_rate}) and using Eq.(\ref{fiducial}) for $r_0$ , 
one finds 

\begin{equation}
\dot{M}_{\rm in} \propto M^{\frac{3}{2}} 
M^{-\frac{2(a+b-1/2)}{5+6a+2b}\left(5/2+2\alpha\right)} \;.
\label{scale_M}
\end{equation}
If $a=0$ and $b=2$ (dust case), it simply reads 
$\dot{M}_{\rm in} \propto M^{\frac{2}{3}(1-\alpha)}$ and with $\alpha=-0.22$
we find a slope of 0.81.
In our model the smallest value $\alpha$ can tend to is $-1/2$  
($\alpha=-1/2$ corresponds to pure radial infall) which 
would give a maximum slope of 1, $\dot{M}_{\rm in} \propto M$ in the
dust dominated regime.

With the purely transit solutions like the previous one, 
the infall rate we calculate corresponds to the rate 
of gas moving towards the protostar. However, there is no
net infall onto the central object in the sense that for
these models, all the matter that falls is deflected in the
outflow. This can be shown by writing 
the continuity equation Eq.(\ref{continuity}) in the 
dimensionless variables and integrating over all space:
\begin{displaymath}
\int_{0+\epsilon}^{\pi/2-\epsilon} 
\left[(1+2\alpha)\mu u_r +\frac{1}{\sin \theta}
\frac{{\rm d}}{{\rm d} \theta}(\mu u_\theta \sin \theta)\right]
\sin \theta {\rm d}\theta 
\end{displaymath}
\begin{equation}
=0
\label{bilan}
\end{equation}
%
The first term corresponds to the total dimensionless mass rate
over the entire domain. With our boundary conditions 
$u_\theta(0)=u_\theta(\pi/2)=0$, and $\mu$ finite, 
the integration of the second term
equals zero. As a consequence, the total mass rate in the model 
is also equal to zero and all the infalling matter in diverted
outward. In order to compare to observations, 
it is possible to compute the outflow to infall rate ratio, 
$f=\dot M_{out}/\dot M_{in}$. For Class 0 and 1 objects, the ratio 
is found to be smaller than unity typically, $f\sim 0.1-0.3$ 
\citep{1996A&A...311..858B,2000prpl.conf..867R}.
In our case, for pure transit models, $f=1$ by construction 
since all the material that is coming in has to be
rejected in the outflow. 

Nevertheless, the inclusion of magnetic field greatly improves the 
properties of the solutions which
now compare well with observational quantities, with fast, 
collimated outflows. 
 
\subsection{Dust case -- a typical solution with net infall}

We now present solutions with a combination of transit and pure infall 
solutions similarly to \citet{Lery02}.
We briefly review the interest and the properties of such solutions.

The method of integration is the same as  previously 
 except that,  here, the domain is separated in two zones. The separation 
occurs at the angle $\theta_s$. From $\theta_0$ to $\theta_s-\epsilon$,
we search for a transit solution, qualitatively identical to 
the one of the previous section. From $\theta_s+\epsilon$ 
to $\pi/2-\epsilon$, we look for a solution that presents
the characteristic of an infall ($v_r<0$) but that is directed
toward the equator and not toward the axis so that the gas
is not deviated. This condition is fulfilled if the 
$\theta-$velocity is positive. At the end of the transit region, 
in $\theta_s$, $v_\theta$ tends to zero and is negative 
(requirement for transit).
To integrate the system in the net infall region, we change the
sign of $v_\theta$, in order to get 
a strictly infalling pattern and use the end
values of the transit solution, as initial conditions for the solution 
between $\theta_s$ and $\pi/2$.

There is a small region around $\theta_s$ which is not 
treated: $\theta=\theta_s$ is a location that cannot be reached
(as for the equator in the pure transit solutions)
and the integration numerically failed before reaching it.
It is also important to mention that the two solutions,
on both sides of $\theta_s$ are mathematically 
independent. The continuity set between the two zones
is in consequence only apparent. 

\begin{figure*}[t]
\begin{center}
   \begin{minipage}[c]{.48\linewidth}
    \includegraphics[clip=, bb=36 145 532 595,width=8cm]{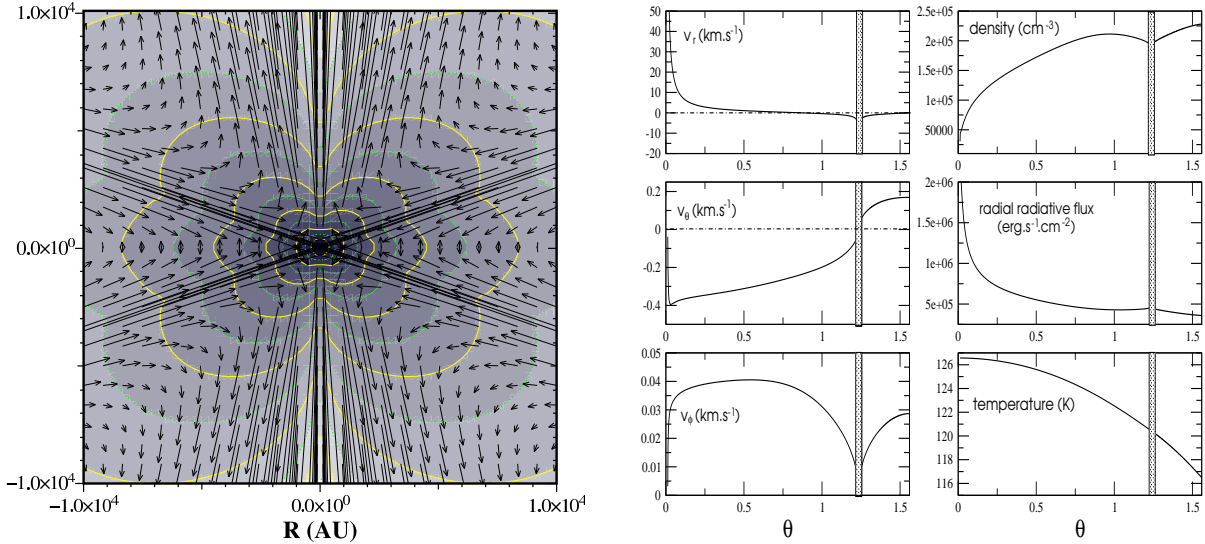}
    \end{minipage} \hfill
   \begin{minipage}[c]{.48\linewidth}
      \includegraphics[height=7.1cm,width=7.5cm]{f7b.eps}
\end{minipage}
\caption{Same as Fig.~\ref{solcirc} but for a solution with net infall.
\label{mhdsol}}
\end{center}
\end{figure*}

A solution  with  the characteristics described above
is shown in Fig.~\ref{mhdsol}. As previously,
we  plot  the density contours and the velocity field (left panel)
as well as the variation of the physical quantities with $\theta$ at a  
given radius (right panel). We are considering the standard case
for which the opacity is dust-dominated. 
 
\label{typicalMHD}
 
The two zones can be distinguished: i) the transit zone,
characterized by the change of the sign of the radial velocity
and by a negative theta-velocity.
(i.e. from $\theta_0=10^{-2}$ to $\theta_s \sim 0.8$~rad), ii) 
a net infalling region
where $v_r < 0$ and $v_{\theta}>0$ in the remaining of the 
domain. In this region, the gas is falling onto the equatorial
plane so it can eventually accrete onto the central protostar.
The untreated region between the two zones has been shadowed.

Fig.~\ref{mhdsol} shows a solution with an opening angle of
$\theta_{\rm open} \sim 30^{\circ}$ but solutions with smaller
opening angles have also been found. The radial velocity for the 
outflow is 20-50~km~s$^{-1}$ and the average density around
$2\times10^{5}$~cm$^{-3}$. 
Again, the solution appears almost spherical: each physical quantity
does not show strong gradient except close to the boundary angles.
Identically with the transit magnetized solution, 
the minimum density and maximum radiative flux is 
in the axis region: this allows the outward acceleration of the gas in
this region and the radial velocity is then also
maximum at the axis.

The $\theta-$velocity tends to zero at the axis and towards $\theta_s$.
These are the boundary conditions we require for transit solution 
and have in consequence to establish a pressure balance between
the transit and net infalling regions. 
Near the equator, $v_\theta$ does not tend to zero. This implies that
the equatorial zone $[\pi/2-\epsilon,\pi/2]$ has to be seen as a
sink that absorbs anything entering it. 
Looking at Eq.(\ref{bilan}) for the net infalling zone (i.e. integrating 
between $\theta_s$ and $\pi/2$), one sees that a non-zero $v_\theta$ 
at one of the boundary ($\pi/2$ in the present case) results in
a non-zero global mass rate in the considered region. This justifies
\emph{a posteriori} terming these solutions ``net infall'' models.
For this category of models, the ratio $f$ is 
then smaller than unity. For the particular
solution shown here, 
$\dot M_{\rm out} \sim 2\times 10^{-5}$ M$_\odot$ yr$^{-1}$
and $\dot M_{\rm in} \sim 2 \dot M_{\rm out}$ resulting in 
$f=\dot M_{\rm out}/\dot M_{\rm in}\sim 0.5$. 
This is a typical value for net infall models, but larger
and smaller values can also be obtained by setting the net
infalling region as a smaller or larger fraction of the 
full model. 

\subsection{Influence of opacity \label{kramer}}

The solutions presented  previously  had a dust-dominated
opacity. This is the situation in the present-day universe
which  is  enriched in heavy elements, molecules and
grains thanks to many generations of stars and stellar nucleosynthesis. 
However, in the primordial universe, star formation occurred in  an almost 
metal-free environment where the chemical composition was 
a mixture of hydrogen (H), helium (He) and very small 
traces of deuterium (D) and light elements (Li, Be, B).  
To understand the physics in the early universe, it is necessary
to have accurate values of its opacities. Many studies 
undertaken on primordial star formation used only  
the species involved
in the formation of molecular hydrogen (and sometimes helium)
\citep{1986PASP...98.1081S,1998ApJ...508..141O,1999ApJ...515..239N} as H$_2$ is
the most abundant molecular species formed under primordial
condition and is also the most effective coolant of the gas 
in the low temperature regime of star formation.
However, HD molecules can also contribute to the cooling 
of the gas when H$_2$ becomes inefficient \citep{1999sf99.proc....6P} 
and in a recent paper, \citet{2005MNRAS.358..614M} show that 
the inclusion of Li in primordial opacity calculation 
lead to significant changes, up to two orders of magnitude for 
$T < 4000$~K and conclude that the influence of Li on the different stages
of population III star formation and evolution could be assessed.
Here, we wish to study the behavior of our models
when the optical   properties of the fluid 
are those of a dust-free environment. 

The use of  
the Kramer's law form for the opacity prevents an explicit
treatment of the cooling that includes all the species
previously mentioned.
However, the Kramer's opacity coefficients ($a,b$) can be interpreted
in term of the physical process enabling the cooling.
The emission rate per unit of volume can be written as 
$\epsilon = 4 \sigma \rho \kappa T^4 \propto\rho^{a+1} T^{b+4}$
\citep{1995MNRAS.276..133B}.
\begin{itemize}
\item $a=0$ corresponds to situations where a single efficient
component (dust or $CO$) dominates the cooling, assuming its 
fraction by mass remains constant during evolution. 
\item On the other hand, $a=1$ gives the density dependence
of the opacity when the cooling is due to two-particle processes
and no coolant is dominant. For such a density dependence, 
the system is expected to be dilute and cold so that the excitation
levels are almost empty.
\item $b$ measures the difference between a black body emission 
($\propto T^4$) and the one considered here. 
In the case of dust, $b=2$ as seen before.
This appears to be an upper limit and $b$ then decreases for 
other types of cooling, in particular $b \in [-5/2,-1]$ for molecular 
cooling in interstellar clouds \citep{1978ApJ...222..881G}. 
\end{itemize}
In order to study the influence of the opacity on the solutions
of our model, we start from the dust case and then increase $a$
from 0 to 1 and decrease $b$ from 2 to 1 so as to leave the
dust-dominated regime. If no dust is present in the 
star formation environment, 
then the main source of opacity would come from 
  molecular line cooling . 
As mentioned previously, 
that case would see $b$ in the range~$[-5/2,-1]$. However, in this
work, $b$ does not take any value smaller than unity. This is due
to the stiffness of the ODE system to solve. 
In fact, to study the influence of opacity on 
the solutions, 
we do not change any other parameters but $a$ and $b$;
for a given set of input parameters, a valid solution (i.e. filling
the whole space and presenting circulation features) appears
to remain valid only for $b$ varying between 2 and 1. Below 1, 
the integration fails before $\theta=\pi/2$ and the other 
input parameters should be slightly adjusted. Doing so would make 
it impossible to distinguish the 
effects due to opacity from those coming from the
new input values. The limit $b=1$ is not strict, we found solutions 
that could still be integrated with $b\lesssim 1$ and others that
would fail earlier. 

By changing
$a$ and $b$ progressively, we depart from the dust-dominated regime. It
is however important to note that we cannot say what type of  
coolant it is we are modelling except that it is less efficient
than dust and results in a smaller opacity. 

The study is made on pure transit 
solutions only and not on net infall ones as the discontinuity
between the two regions of the latter was more difficult to handle.

\begin{figure}[ht]
\begin{center}
\includegraphics[bb=50 50  730 580, width=8cm]{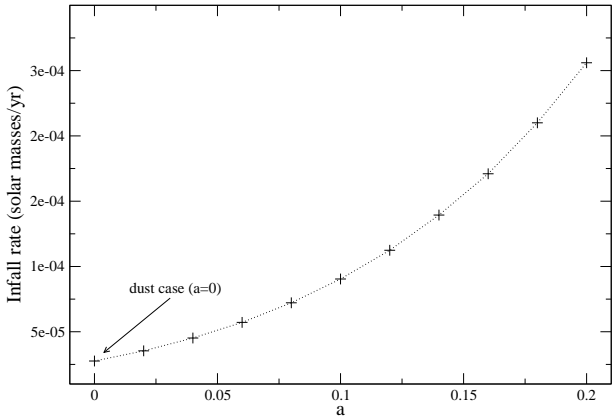}
\caption{Influence of Kramer's opacity parameter $a$ on the infall rate.}
\label{influence_a}
\end{center}
\end{figure}

\begin{figure}[ht]
\begin{center}
\includegraphics[bb=50 50  730 580, width=8cm]{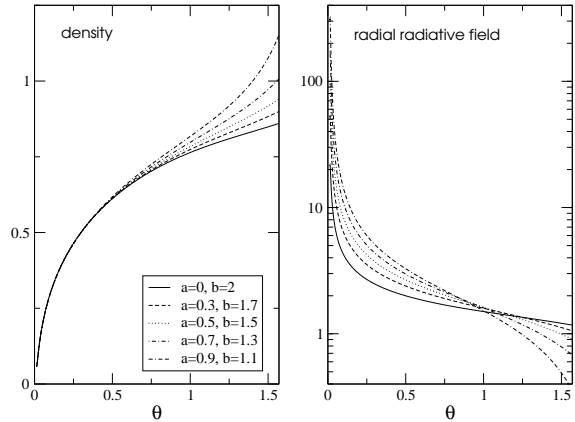}
\caption{Influence of the opacity parameters $a$ and $b$ on the
dimensionless density and radial radiative flux.}
\label{graph_opacity}
\end{center}
\end{figure}

The results are shown in Fig.~\ref{influence_a} and 
Fig.~\ref{graph_opacity}.
The infall rate is calculated in the same way as in Sect.~\ref{dust}
and is plotted in Fig.~\ref{influence_a} as a function of the 
Kramer's opacity parameter $a$ ($b$ stays equal to 2). 
The infall rate increases as one leaves the dust opacity regime.
Very simply, when the opacity decreases and the medium becomes
more transparent, the photons escape more easily and are less
efficient in slowing the infalling gas.

In Fig.~\ref{graph_opacity}, the left panel
represents the density for different sets ($a,b$). The radial radiative
fluxes are plotted on the right panel for the same sets of parameters.
Furthermore, the quantities plotted are dimensionless
(i.e., $\mu(\theta)$ and $f_r(\theta)$)   
because as $a$ and $b$ are modified, the fiducial scale
$r_0$ changes (see Eq.(\ref{fiducial})): 
the influence of the opacity on the morphology of
the solutions would be more difficult to visualize as they would not
start from the same point.

The dimensionless density increases
in the equatorial region as one leaves the dust-dominated 
regime. As it is the dimensionless density, this means that
as $a$ increases and $b$ decreases, the ratio between the density
at the equator and at the rotational axis increases.
The radial radiative flux shows the opposite behavior: the flux is relatively
smaller at the equator than in the axial region.

It is not possible to generalize this behavior as not all 
the solutions that were found behave like that. For example, the 
morphology of the solution presented in Sect.\ref{dust} was almost 
unaffected by the change of opacity. The only difference was
observed for its dimensional quantities that would scale 
differently and, as a result, give a higher infall rate when 
dust is not the dominant coolant (see Fig.~\ref{influence_a}).
However, it is of interest that such solutions exist: the higher 
density in the equatorial region along with the weaker radiative
field in this region could lead to the accumulation of 
matter in the equatorial plane (i.e., to a more massive accretion 
disk) and may ultimately lead to the formation of more massive star
for   dust-free opacity regimes 
(such as the ones of the primordial universe).  

\section{General properties of MHD models \label{Gprop}}
The main feature of the model is the production of a heated 
pressure-driven outflow with magneto-centrifugal acceleration and collimation.
An evacuated region exists near the axis of rotation where the 
outflow is produced and the density systematically increases with 
the angle from the axis. The streamlines passing close to
the rotational axis are also the ones closest to the central object
(see Fig.~\ref{Bernoulli}, upper left panel). The 
gas on these streamlines transits deeper in the gravitational well
and is also the most vigorously heated. As a consequence, the radial
velocity of the outflow is always maximum near the axis and decreases
with increasing poloidal angle $\theta$.

\subsection{Two families of solutions}

\begin{figure}[ht]
\begin{center}
\includegraphics[bb=50 50  730 580, width=8cm]{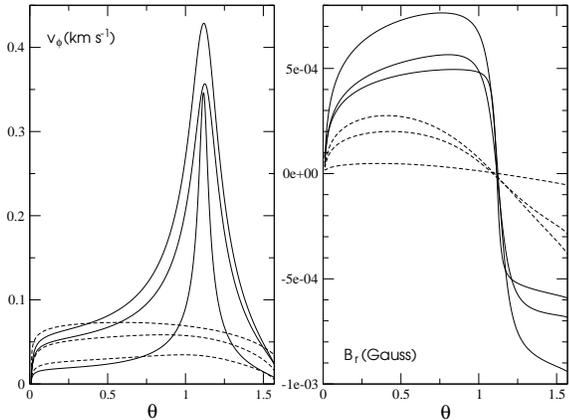}
\caption{The rotational velocity (left) and radial magnetic field (right)
plotted as a function of $\theta$. The continuous and dotted lines
correspond to the two types of solutions. For clarity, we only show
solutions with almost the same turning angle, but others were found.
\label{family}}
\end{center}
\end{figure}

 All the solutions that we have found have the same qualitative properties, 
except around  
their turning point. In this region, two distinct families of solutions 
appear as shown in Fig.~\ref{family}.
The rotational velocity $v_\phi$ 
(left panel) and the radial magnetic field $B_r$ (right panel)
are plotted as a function of the poloidal angle $\theta$ for different
solutions. 

The solutions 
fall into one of the two categories: i)~slow rotators
where the rotational
velocity is small and almost constant over the entire domain (dotted line) 
and ii)~fast rotators 
where $v_\phi$ increases and decreases with a strong gradient
around the turning point of the solution (solid line). 
At its maximum, $v_\phi$ peaks
at about four times the $v_\phi$ value of the other solutions.
Because we have $B_\phi \propto v_\phi$, the solutions that peak 
at the turning point also have a much stronger toroidal field there and
as a consequence a larger pinch force towards the rotational axis.

The radial magnetic field (Fig.~\ref{family}, right panel) 
changes sign at the turning point 
of the solution ($B_r \propto v_r$) and it appears that the 
fast rotating solutions (solid lines) have a substantially higher 
radial magnetic field on the overall domain than the slow rotating
models (from 2 to 16 times higher for the solutions shown).

The other variables (not shown here: density, temperature, etc...) 
do not seem to be affected and no distinction can be made between 
the two types of solutions when looking at these quantities. 
  No morphological differences appear between 
the two types of models and the density profiles all indicate the 
characteristic oblate shape of magnetized solutions. 
Rotation and magnetic field seem coupled and
the solutions with reasonable dimensioned quantities are
in one of the two following categories: i) slow rotating and poorly magnetized
or ii) faster rotating and highly magnetized.

\subsection{Energetics of the solutions}
In order to compare and contrast the two families of solutions, the evolution 
of the energy along a streamline is considered. We call Family~1/Family~2
the slow/fast rotating weakly/strongly magnetized 
family of solutions found  in the previous section.

Streamlines follow
${\rm d}r/(r {\rm d}\theta)=u_r/u_\theta$ in the poloidal plane. 
It is convenient to look at the local specific energy components
normalized to the local specific gravitational energy (to have a dimensionless
quantity) and to define their sum as
\begin{equation}
Be(r,\theta)\equiv -\frac{1}{E_{\rm grav}}\left(E_k+E_{\rm grav}
+\frac{p}{\rho}+E_{\rm magn}\right)
\end{equation}
where $E_k$ is the specific kinetic energy, 
$E_{\rm grav}=-GM/r$ the specific gravitational
potential energy, and $E_{\rm magn}=B^2/8\pi\rho$ the specific 
magnetic energy. Using the self-similar form for the quantities,
$Be$ becomes
\begin{equation}
Be(\theta)=\frac{1}{2}\left(u_r^2+u_\theta^2+u_\phi^2\right)-1+\Theta
+\frac{1}{8\pi\mu}\left(\frac{u_p^2}{y_p^2}+\frac{u_\phi^2}{y_\phi^2}\right)
\end{equation}
Note that in the hydrodynamical case, the Bernoulli theorem implies that 
$Be$ is constant along a streamline.

\begin{figure}[t]
\begin{center}
\includegraphics[bb=50 50  730 580, width=8cm]{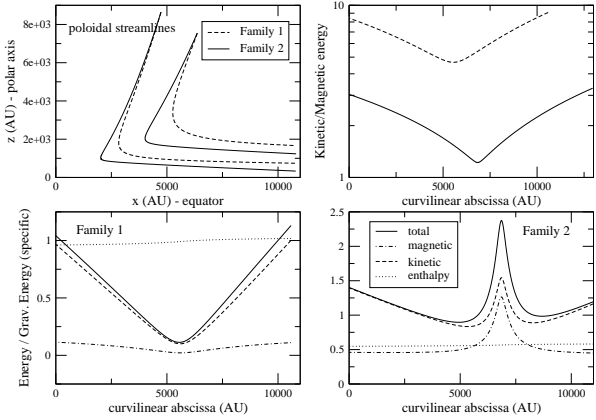}
\caption{Family 1 and 2 refer respectively to the slow and fast rotating
solutions found previously.
\emph{Upper left}: poloidal streamlines. 
\emph{Upper right}: ratio of the kinetic to the magnetic energy
as a function of the curvilinear abscissa $s$.
$s=0$ is chosen at the infalling end of the streamline. 
\emph{Lower left}: energetics of a solution 
along a streamline for a solution from Family 1. The specific
energy components have been normalized to the gravitational potential
energy. \emph{Lower right}: 
same as lower left for a solution from Family 2. \label{Bernoulli}}
\end{center}
\end{figure}

In Fig.~\ref{Bernoulli}, upper left panel, streamlines are plotted 
for two solutions, one from each family. The central object is located 
at the origin. It is interesting to note that for streamlines integrated from 
the same point (located in the outflowing region), 
the solution from Family 2 transits closer to the central object,
deeper in the gravitational well than the one from Family 1. This is 
a general trend that has been observed for many solutions.
The three other panels in Fig.~\ref{Bernoulli} show energy components
plotted along these two streamlines.

On lower left and right panels the different specific energy
components as a function of the position along the streamline of the 
solution from Family 1 and 2 are plotted.
The origin of the curvilinear abscissa $s$ is chosen at the ``infalling
end'' of the streamline. 
 
Far from the turning point, the main contribution to the 
kinetic energy ($E_k$) comes from the radial velocity $v_r$ ($v_\theta$ and 
$v_\phi$ have much smaller values).
The absolute value of $v_r$ rapidly decreases when 
approaching the turning point
(See Fig.~\ref{solcirc}) and so does  $E_k$ (dashed line),
for both Family 1 and 2 solutions.
Near the turning point, however, the contribution of $v_{\theta,\phi}$
to $E_k$ becomes dominant: i) for the slow rotating solutions 
(Family 1 -- lower left panel), 
$v_\phi$ does not change much on the domain and $E_k$
reaches a minimum at the turning point, ii) but for the fast rotators (lower
right), $v_\phi$ quickly increases at the turning point and $E_k$ reaches
a local maximum.
The magnetic energy qualitatively follows the same behavior
as $B_{r,\theta} \propto v_{r,\theta}$ and $B_\phi \propto v_\phi$.

Finally, the ratio of the kinetic to the magnetic energy is plotted
on the upper right panel in Fig.~\ref{Bernoulli}, 
again for the two types of solutions.
Qualitatively, they behave alike, and $E_k/E_{\rm magn}$ is in 
both case minimum at the turning point. This conforms with intuition 
as one expects a larger influence of the magnetic field near the 
central protostar. 
However, at the turning point, 
$E_{\rm magn}$ is 5 times smaller than $E_k$ for the slow
rotating solution while it is comparable to  $E_k$ for the 
fast rotating model. For clarity only two solutions have been plotted here,
but it has been found a general behaviour that 
$(E_k/E_{\rm magn})^{\rm Family 2 }_{\rm turn.} \sim 1$. Furthermore,
the same ratio for Family 1 solutions has been found many times much
greater (5-100) than unity.

In conclusion, the behaviour of the two types of solutions is
similar far from their turning point, but then differs significantly:
for the slow rotating models, the magnetic energy is always much smaller than
the kinetic energy whereas they become comparable
for the fast-rotating solutions. The magnetic field appears then to be
responsible for a closer transit around the center of the latter ones.
Despite this difference between the two families of solutions, 
the properties (velocity, density) of the outflows they power are similar.

\section{Discussion \label{discussion}}

\subsection{Behavior at large distance}

After the main early phase where infall dominates the dynamics,
the protostar deposits linear and angular momentum and mechanical energy into
its surroundings through its jets and molecular outflow. It is therefore
important to discuss the behavior of the outflows at large
distance. In the case of low mass stars, 
the accretion-ejection engine should dominate the 
global dynamics and produce relatively weak molecular outflows. 
In that case, the present model may apply only for a short stage
at the earliest phases of low mass star formation. 
On the other hand, if the physics of the outflow formation remains the
same regardless the mass of the protostar,
the dynamics of massive objects should be dominated by 
the transit and should have strong and heavy molecular outflows,
according to the present model. In that case, the transit model plays
a major role in the formation of massive stars.
Moreover, the transit model brings an interesting new
feature for the interaction  between molecular outflows
and jets. Indeed, a lot of gas is brought to the axial region
that is not at rest and is already stratified when the atomic
jet starts propagating away from the protostar. It will drastically 
change the jet propagation and stability properties.
This demonstrate the importance of understanding of the history
of outflows in order to model correctly the behavior of
jets. 

Also, according to our model, at larger distance, 
$\rho \propto r^{2\alpha-1/2}$, $B\propto r^{\alpha-3/4}$, 
and $v\propto r^{-1/2}$. The magnetic field and the density 
decrease faster than the velocity. For example for $\alpha=-1/4$, 
$\rho \propto r^{-1}$, $B\propto r^{-1}$, and $v\propto r^{-1/2}$,
and the outflow becomes purely ballistic \citep{Lery02}. 
Only large bow-shocks due to the long time-scale
variations of the source should survive.
Ultimately, the combined effects of the reduction of the 
ionization and temperature in the giant 
bows, and of the lack of ambient medium, may make the outflows 
fade out and finally become very difficult to observe on distances of 3 to 10 pc.
However, the momentum that they may transfer to the ambient medium
could help to maintain turbulence on large scales.

\subsection{The outflow evolution}
Of our steady-state model, any temporal evolution is beyond the
scope. However, we may speculate on the evolution of YSOs by
considering the slow quasi-steady evolution of our model.
With the new insight given by the model, we suggest some physical 
justifications for the early evolutionary sequence of both low and high mass YSOs 
which is usually defined on the basis of an empirical scheme only.

During the Class 0 stage, most of the mass of the system is still
located in the infalling envelope. The source is totally embedded and
the transit of matter around the center object can start during the 
formation of the stellar core. This gives rise to a fast, powerful
and collimated molecular outflow. However during this early stage, the 
central object may not have already developed the jet that is commonly
thought to drive the molecular outflow.

When reaching the early Class 1 stage, the central object is surrounded by
both a disk and a diffuse circumstellar envelope. A jet eventually 
emerges from the newly formed inner accretion disk as the transit
continues. The jet may then pressurize and entrain 
the material in the axial region, pushing the transit flow away from
the axis, thus reducing the molecular outflow velocity and increasing
its diameter. Therefore, molecular outflows will appear to be less
powerful and less collimated with time.

For low mass protostars, in the latest stages of the pre-stellar evolution, 
the central object should not be embedded in an envelope and our model 
cannot apply. The molecular outflow continues to 
spread out and slow down and is eventually overtaken by the central jet still
fed by the accretion disk. 
In the case of high mass stars, the transit features may remain in action 
until the central star forms and reach the main sequence. 
Either the increasing opening angle of the 
molecular outflow in the low mass case, or the ignition of the central object
may ultimately stop the accretion by suppressing
any contact between the reservoir formed by the molecular cloud and 
the accretion disk or the star.

\section{Summary}

In this paper, we have presented 
a MHD model that applies to infall and molecular
outflows around young stellar objects. 
The model is based on the radial self-similarity
assumption applied to the basic equations of ideal, 
axisymmetric and steady-state MHD, including Poynting flux. Instead of the
usual mechanisms invoked for the origin of molecular outflows (underlying
jet or wind), the outflow is powered by the infalling matter through 
a heated quadrupolar transit pattern around the central object.

Solutions without magnetic field have been found and they appear
almost spherical (quantities do not vary much with the poloidal angle). 
Although they do 
not compare well with observation ranges (too dense and too slow outflow), 
they indicate that thermodynamics
is a sufficient engine to generate an outflow. They could apply to 
star formation in the primordial universe or 
in the very early stage of star formation where magnetic fields may not
be  dynamically important (however, in the latter case, the point
mass gravity field should be replaced by self-gravity). 

The magnetized solutions show dynamically significant density gradients 
in the axial region, precisely where the radial velocity and 
collimation are the largest. Their radiative field is also highly 
anisotropic and the highest at the rotational axis. It contributes to
the ejection of the gas.
Quantitatively, the pure transit solutions 
compare well with observations of outflows: velocities
of a few tens of km~s$^{-1}$ and typical density of $10^{4-5}$~cm$^{-3}$.

However, for these models all the infalling
gas is deviated into the outflow -- there is no net infall. 
It is possible to obtain, from the 
same set of equations, a purely infalling region around the equatorial
plane. The gas in this region is not deviated and will likely end up
on the accretion disk around the protostar.
When changing the scaling and putting a more massive central 
object, the infall rate also increases. This indicates that
our model favors the large accretion rate scenario to form
massive stars.

The influence of the opacity on the transit solutions has been
studied. When leaving the dust dominated regime, the fiducial
scale of the model decreases which results in an increase of
the infall rate. Furthermore, the morphology of some solutions
is also affected: the density ratio between the equator and the axis
increases while the radial radiative flux ratio decreases. As
a consequence, when dust does not dominate the cooling, such as in
the primordial universe, matter could be more easily accumulated in the
equatorial region   and form more massive stars .

We suggest that molecular outflows are dominated by the
global transit of material around the protostar (except for a thin layer
surrounding the central jet -- if present, 
where the dynamics is governed by entrainment) and that the same 
process occurs from low to high mass forming stars.
The present work also suggests
that radiative heating and magnetic field may ultimately be
the main energy sources driving outflows during star formation, 
at the expense of gravity and rotation.

Finally, although the details of the outflow mechanisms may be peculiar
to individual objects we believe the infall-outflow circulation
arises naturally given accretion, and thus could also
be present in other astronomical objects such as active galaxies
and around suitably placed compact objects, such as neutron stars. 
\acknowledgments
The authors are 
particularly grateful to Prof. R.N.~Henriksen and Dr. J.~Fiege
for their contribution to the field and their particular support 
and discussions. C.C. would also like to thank S.~Leygnac
and D.~Maurin for useful discussions.
This work was carried out as part of the CosmoGrid project, funded
under the Program for Research in Third Level Institutions (PRTLI)
administered by the Irish Higher Education Authority under the National
Development Plan and with partial support from the European
Regional Development Fund.
The present work was supported in part by the European CommunityÕs 
Marie Curie Actions - Human resource and mobility within the JETSET 
network under  contract MRTN-CT-2004 005592.


\end{document}